\documentclass[aip,reprint]{revtex4-2}

\usepackage{color}
\usepackage{graphicx}
\usepackage{amsmath}
\usepackage{subcaption}

\usepackage{natbib}

\begin{document}
\title{Magnetized ICF implosions: ignition at low laser energy using designs with more ablator mass remaining}

\author{C. A. Walsh}
\email{walsh34@llnl.gov}
\affiliation{Lawrence Livermore National Laboratory, 7000 East Avenue, Livermore, CA 94550, USA}
\author{S. T. O'Neill}
\affiliation{University of York, Heslington, York YO10 5DD, UK}
\author{D. J. Strozzi}
\affiliation{Lawrence Livermore National Laboratory, 7000 East Avenue, Livermore, CA 94550, USA}

\author{L. S. Leal}
\affiliation{Lawrence Livermore National Laboratory, 7000 East Avenue, Livermore, CA 94550, USA}
\author{R. Spiers}
\affiliation{University of Delaware, 210 S College Ave, Newark, DE 19716, USA}

\author{A. J. Crilly}
\affiliation{Imperial College, Exhibition Rd, South Kensington, London SW7 2AZ, UK}
\author{B. Pollock}
\affiliation{Lawrence Livermore National Laboratory, 7000 East Avenue, Livermore, CA 94550, USA}
\author{H. Sio}
\affiliation{Lawrence Livermore National Laboratory, 7000 East Avenue, Livermore, CA 94550, USA}
\author{B. Hammel}
\affiliation{Lawrence Livermore National Laboratory, 7000 East Avenue, Livermore, CA 94550, USA}
\author{B. Z. Djordjevi\'c}
\affiliation{Lawrence Livermore National Laboratory, 7000 East Avenue, Livermore, CA 94550, USA}
\author{O. Hurricane}
\affiliation{Lawrence Livermore National Laboratory, 7000 East Avenue, Livermore, CA 94550, USA}

\author{J. P. Chittenden}
\affiliation{Imperial College, Exhibition Rd, South Kensington, London SW7 2AZ, UK}
\author{J. D. Moody}
\affiliation{Lawrence Livermore National Laboratory, 7000 East Avenue, Livermore, CA 94550, USA}

\date{\today}

\begin{abstract}
This paper is the first work to redesign a spherical ICF implosion to best utilize the benefits of applying an external magnetic field. The sub-ignition experiment N170601 is taken as the baseline design, which used 1.57MJ of laser energy. The optimum magnetized design benefits from increasing the shell thickness by 14$\mu$m and decreasing the ice thickness by 18$\mu$m, resulting in a neutron yield of 8.9$\times$10$^{17}$. This is 34$\times$ greater than the unmagnetized simulation of the same design, and 18.5$\times$ the greatest unmagnetized simulation across all designs simulated. The resultant implosion velocity for the magnetized design is lower, which would also reduce ablation front instability growth. This design was found by using a simplified 1D magnetization model, then validated against full 2D extended-MHD capsule simulations with radiation asymmetries applied to correct the shape.
\end{abstract}
\maketitle

Magnetic fields can be used to increase performance of inertial confinement fusion (ICF) implosions through suppression of thermal conduction\cite{hohenbergerInertialConfinementFusion2012a,walshMagnetizedICFImplosions2022,slutzHighGainMagnetizedInertial2012a}, magnetization of $\alpha$-particles\cite{oneill2025} and suppression of instability growth\cite{chandrasekharHydrodynamicHydromagneticStability1962,srinivasanMitigatingEffectMagnetic2013,walshMagnetizedAblativeRayleighTaylor2022}. The first magnetized spherical implosions used an 8T axial field on direct-drive experiments on the OMEGA Laser Facility, resulting in a 30\% increase in the yield \cite{changFusionYieldEnhancement2011}. More recently a 26T field has been applied to indirect-drive implosions on the National Ignition Facility (NIF), improving the yield by a factor of 3 \cite{moodyIncreasedIonTemperature2022}.

While experimental results have been promising, it is thought that the 26T applied on NIF is already reaching the maximum benefit due to thermal conduction suppression\cite{sioDiagnosingPlasmaMagnetization2021,walshMagnetizedICFImplosions2022}. Theoretical scalings also suggest that magnetic fields benefit low-temperature hot-spots more than high-temperature \cite{walshMagnetizedICFImplosions2022}. Questions naturally arise: is the maximum yield benefit from a magnetic field 3$\times$? Will magnetic fields actually be useful in igniting systems, where the hot-spots are already high temperature?

All current computational work has taken an established unmagnetized spherical implosion design and added a magnetic field \cite{strozzi2024,perkinsPotentialImposedMagnetic2017}. This paper is the first research to search for a more appropriate target that utilizes magnetization effects in the design process. Section \ref{sec:1D} uses a simplified 1D magnetization model for ease of scanning target specifications (shell and ice thickness). A new optimal capsule design with thicker shell and thinner ice for use with applied fields is suggested, resulting in a lower implosion velocity and more mass remaining during hot-spot burn. Section \ref{sec:2D} then validates this design with 2D extended-MHD simulations. By correcting the inherent magnetization asymmetry with radiation drive asymmetries, more than half of the idealized 1D magnetization model yield can be achieved.

A magnetic field modifies the electron heat-flow in a plasma to be anisotropic \cite{braginskiiTransportProcessesPlasma1965}:

\begin{equation}
	\underline{q}_e = -\kappa_{\parallel} \nabla_{\parallel}T_e -\kappa_{\bot} \nabla_{\bot}T_e - \kappa_{\wedge} \underline{\hat{b}} \times \nabla T_e  \label{eq:magheatflow}
\end{equation}
Where $\kappa_\parallel$ is the thermal conductivity along field lines, $\kappa_{\bot}$ is the thermal conductivity perpendicular to field lines, and $\kappa_{\wedge}$ is the Righi-Leduc coefficient. $\kappa_{\parallel}$ remains unchanged due to magnetization, while $\kappa_{\bot}$ decreases with the electron Hall parameter, $\omega_e \tau_e$. The Righi-Leduc coefficient peaks for $\omega_e \tau_e\approx 0.1-1.0$, but is found to be insignificant in these pre-magnetized systems\cite{walshExtendedMagnetohydrodynamicEffects2018}.

$\alpha$-particles produced through DT fusion reactions can also be trapped by magnetization. A 3.5MeV $\alpha$-particle with velocity orthogonal to a 10kT magnetic field is confined to a radius of 27$\mu m$, which is similar to a typical hot-spot radius on the NIF. 

The nominal NIF design chosen for this study is N170601 \cite{clark2019}, which used 1.57MJ of laser energy (substantially lower than the current maximum energy, which is now routinely above 2.0MJ \cite{PhysRevLett.132.065102}). This was the first experiment to exceed $10^{16}$ neutron yield, giving 1.47$\times 10^{16}$; this is considered a sub-igniting capsule, with an energy gain of 0.027. The target consists of a 70$\mu$m thick high-density-carbon ablator and 56$\mu$m of DT ice. The total capsule radius is 980$\mu$m, with the central region being DT vapor. 

The Gorgon extended-magnetohydrodynamics (MHD) code \cite{ciardiEvolutionMagneticTower2007,chittendenRecentAdvancesMagnetohydrodynamic2009,walshSelfGeneratedMagneticFields2017} is utilized in this work. Thermal conduction in Gorgon is fully anisotropic \cite{sharmaPreservingMonotonicityAnisotropic2007}, as per equation \ref{eq:magheatflow}. $\alpha$-particles are propagated using a monte-carlo method \cite{tongAlphaHeatingPowerBalance}, with helical trajectories to account for magnetization \cite{oneill2025}. Magnetic transport includes the bulk advection with the plasma, Nernst advection down temperature gradients, resistive diffusion and Biermann Battery \cite{walshExtendedmagnetohydrodynamicsDensePlasmas2020}. Improved magnetic transport coefficients are used \cite{sadlerSymmetricSetTransport2021,daviesTransportCoefficientsMagneticfield2021}, which were found to reduce magnetic field twisting in pre-magnetized simulations \cite{2021}. Gorgon compares favourably with cylindrical magnetic flux compression experiments on OMEGA, increasing confidence in its magnetic transport treatment for pre-magnetized implosions \cite{10.1088/1361-6587/ac3f25}.  New magnetized flux limiters have been implemented into Gorgon \cite{walshKineticCorrectionsHeatflow2024}; these more closely resemble the heat-flow and Nernst advection predicted by Vlasov-Fokker-Plank simulations, although  the results here show little sensitivity to flux limiter value.

Simulations in this paper only take into account the capsule, with no effort made to simulate the hohlraum. A frequency dependent radiation source is used for that purpose, with the radiation drive assumed constant across all designs simulated. The radiation transport algorithm in Gorgon is a P1/3 automatic flux limiting approach \cite{jenningsRadiationTransportEffects2006,mcglincheyDiagnosticSignaturesPerformance2018}.

\section{1D Magnetization Model \label{sec:1D}}

\begin{figure}
	\centering
	\begin{subfigure}{\linewidth}
		\centering
		\includegraphics[scale=1.0]{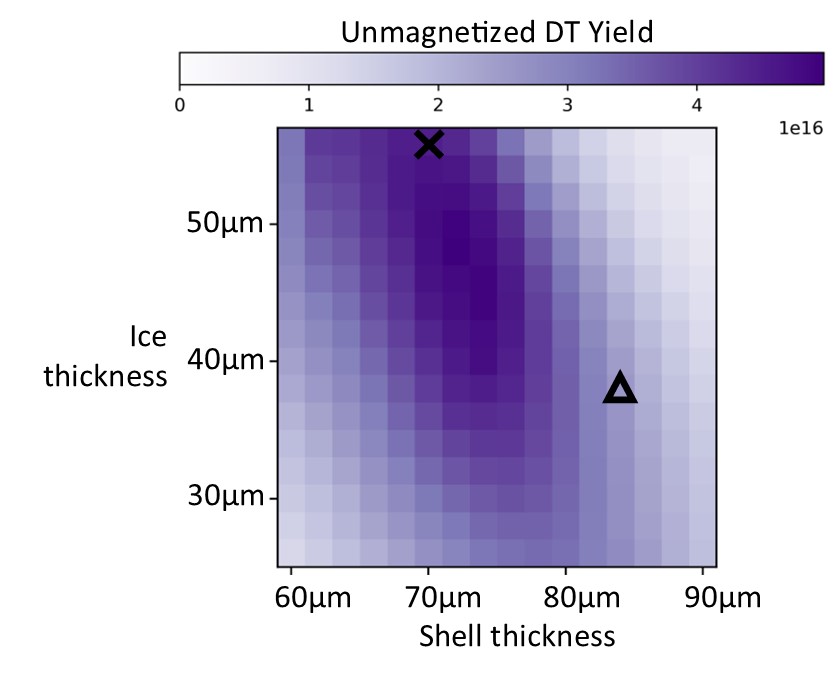}
		\caption{\label{fig:shicethick_unmag}}
	\end{subfigure}
	
	\begin{subfigure}{\linewidth}
		\centering
		\includegraphics[scale=1.0]{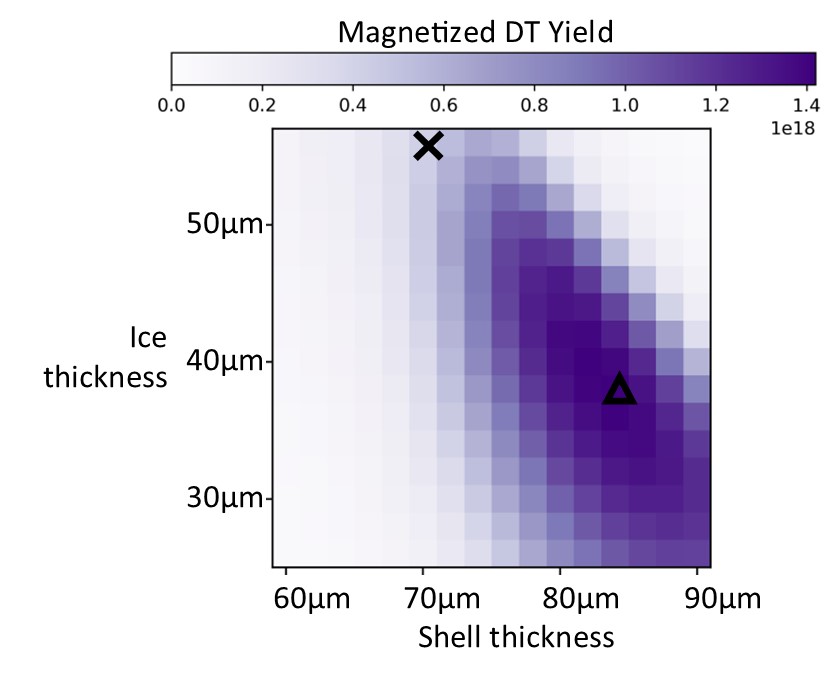}
		\caption{\label{fig:shicethick_mag}}
	\end{subfigure}
	\caption{ \label{fig:shicethick} Neutron yield of 1D simulations varying DT ice thickness and HDC shell thickness. (a) uses unmagnetized simulations. (b) uses a magnetization approximation where the thermal conduction in the fuel is reduced by 2/3 and the $\alpha$-transport is suppressed in 2 of the 3 dimensions. Black 'x' denotes the nominal design point N170601. The triangle shows the magnetization optimum.    }	
\end{figure}

Imposition of a magnetic field naturally adds asymmetries to the stagnation phase of an implosion, which typically requires a 2D description to capture the differing hot-spot quantities at the pole compared to the waist \cite{perkinsPotentialImposedMagnetic2017,walshPerturbationModificationsPremagnetisation2019}. The anisotropic effect of heat-flow \cite{walshMagnetizedICFImplosions2022,walshMagnetizedDirectlydrivenICF2020}, $\alpha$-heating \cite{oneill2025} and perturbation growth \cite{srinivasanMitigatingEffectMagnetic2013,walshMagnetizedAblativeRayleighTaylor2022,walsh2024} are all well studied. In addition, the magnetic field transport can only be captured in 2D, with the field primarily compressed at the waist. 

Nonetheless, a 1D approximation to magnetization effects is desirable for investigating the optimal design once a magnetic field is applied. 2D MHD simulations have shown that the magnetized capsule yield is highly dependent on hot-spot shape \cite{walshMagnetizedICFImplosions2022}; scanning the potential target designs as well as tuning for shape is a computationally expensive task. 1D simulations, however, allow for much faster computation of these effects and do not require shape modifications.  

Recent work found that the effect of magnetization on heat-flow can be approximated using an effective thermal conductivity, $\kappa_{eff}$ \cite{walshMagnetizedICFImplosions2022}. For an axial field applied to a spherical hot-spot, the heat-flow can only be suppressed in 2 of the 3 directions:

\begin{equation}
	\kappa_{eff} = \frac{1}{3} + \frac{2}{3} \frac{\kappa_\perp}{\kappa_\parallel}
\end{equation}

For a fully magnetized spherical implosion with an axial field, $\kappa_{eff}=1/3$. Other topologies can allow for lower effective thermal conductivities (such as a mirror field, which can give $\kappa_{eff}=<0.2$) \cite{walsh2025}. Non-axial fields are not considered in this study, but could be utilized to enhance magnetization benefits\cite{walsh2025}. 
 
Here a simplified magnetization model is first used to quickly scan the target parameter space. Then, the proposed new design will be investigated with 2D extended-MHD simulations in section \ref{sec:2D}, giving broadly comparable results. 

No magnetic field is imposed in the simplified 1D magnetization model. Instead, the magnetic field is assumed to be sufficiently strong to suppress all thermal conduction in the fuel in 2 of the 3 dimensions (i.e. the thermal conductivity is suppressed to 1/3 of its nominal value), The same approach is taken for the transport of $\alpha$-particles. The $\alpha$-particle transport in Gorgon is always conducted on a 3D grid, with coupling then mapped onto the 1D spherical grid \cite{tongAlphaHeatingPowerBalance}. This makes implementation of the approximated magnetization effect simple; the $\alpha$-particles are only allowed to propagate along the axial direction.

No attempt is made here to incorporate the effect of magnetic fields on perturbation growth, which is thought to be an additional benefit \cite{chandrasekharHydrodynamicHydromagneticStability1962,srinivasanMitigatingEffectMagnetic2013,walshMagnetizedAblativeRayleighTaylor2022}. 

All simulations here use 1$\mu$m radial resolution. The magnetization model is switched on once the first shock converges on the implosion axis. The radiation drive is unchanged between all simulations; future work will look to also optimize the radiation drive to best utilize magnetization effects.

First a series of 256 unmagnetized 1D simulations are executed, scanning the shell and ice thickness. Total capsule radius has also been investigated but found to have less sensitivity, so the results are not included here.  The unmagnetized yields are shown in figure \ref{fig:shicethick_unmag}, with the black cross denoting the nominal N170601 design, which is found to be within 10\% of the optimal unmagnetized design calculated here (neutron yield = 5.0$\times$10$^{16}$).

Next, the same process is conducted using the 1D model approximating the impact of reduced thermal conductivity and $\alpha$ transport.These yields are plotted in figure \ref{fig:shicethick_mag}. The black triangle shows the new optimum that has been found, which gives a yield of 1.4$\times 10^{18}$, a factor of 30 greater than the peak yield observed in the unmagnetized parameter scan. These scans suggest increasing the shell thickness by 14$\mu m$ from 70$\mu m$ to 84$\mu m$ and reducing the ice thickness from 56$\mu m$ to 38$\mu m$. 

When $\alpha$-heating is turned off for the new magnetized design the yield drops to 1.3$\times 10^{16}$, i.e. the yield enhancement due to $\alpha$-heating is greater than 100$\times$. This is substantially larger than the $<$30$\times$ yield amplification from $\alpha$-heating estimated for recent implosions on NIF that exceeded 1MJ of yield and passed the Lawson criteria for ignition \cite{abu-shwarebLawsonCriterionIgnition2022}.

\begin{figure}
	\centering
	\centering
	\includegraphics[scale=1.0]{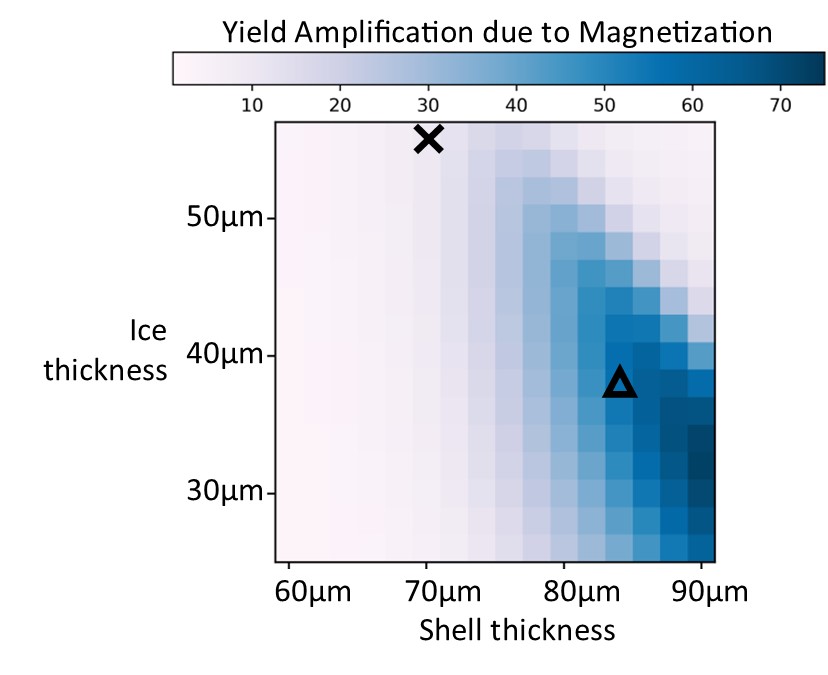}\caption{\label{fig:shicethick_Yamp} Ratio of the 1D magnetized yields (Figure \ref{fig:shicethick_mag}) to unmagnetized yields (Figure \ref{fig:shicethick_unmag}).  }	 
\end{figure}

Figure \ref{fig:shicethick_Yamp} plots the ratio of magnetized yield to unmagnetized yield at each point in parameter space. The magnetic field is found to increase the yield of the new design by 50 times, with that number still increasing as the shell is thickened further and the ice thinned. Up to a factor of 70 was calculated in this scan, although the absolute magnetized yield begins to decrease if the shell is made too thick or the ice made too thin.

\begin{figure}
	\centering
	\centering
	\includegraphics[scale=0.5]{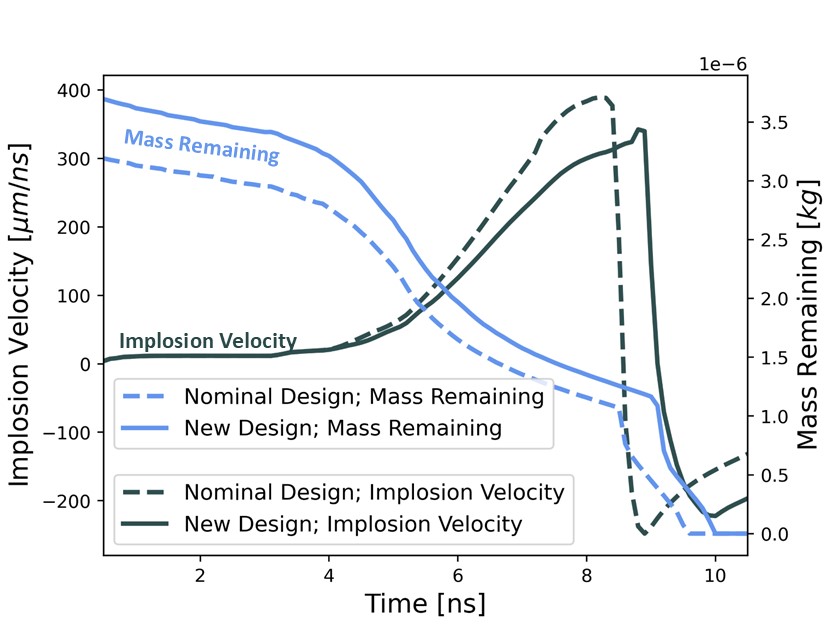}\caption{\label{fig:Mass_Remaining} Mass remaining and implosion velocity over time for the unmagnetized 1D simulations of the nominal N170601 design (dashed line) and the new design (solid line).}	 
\end{figure}

To understand this result further, the nominal design (black cross) is compared with the new design (black triangle) using the unmagnetized 1D simulations. The implosion velocity, defined as the inward velocity of the peak shell density, is tracked over time in both cases and plotted in figure \ref{fig:Mass_Remaining}. Due to the additional shell mass, the new design results in a slower implosion. Note that this would also be beneficial for ablation front stability, although this effect is not captured in the simulations due to no target defects being included here. 

The ablator mass remaining is also calculated for the two cases. This is defined as the total ablator mass that is imploding (rather than exploding). This is also plotted in figure \ref{fig:Mass_Remaining} for both the nominal and new designs. Naturally, there is consistently more mass when the ablator is made thicker. The thicker shell is also expected to enhance implosion stability, as the perturbation must grow more before it can puncture the shell. 

Additional insight can be obtained by comparing the unmagnetized optimum design with the magnetized optimum without $\alpha$-heating included. The neutron yield only increases from 1.0$\times$10$^{16}$ to 1.3$\times$10$^{16}$, while the burn averaged ion temperature increases from 3900eV to 4700eV. However, the amplification of yield by $\alpha$-heating increases from 5$\times$ for the unmagnetized optimum to $>$100$\times$ for the magnetized optimum.

The increased ablator mass remaining during hot-spot formation for the new design results in longer confinement times for the fuel, which is particularly beneficial once $\alpha$-heating is substantial \cite{maclaren2021}. Other capsule designs have tried to utilize this effect, such as the pushered single shell targets that proposes a dense layer of high-Z material on the inside surface of the shell \cite{maclaren2021}. The results in this paper suggest that this can be accomplished by magnetization without the drawbacks of additional unstable interfaces and the potential for high-Z material to mix into the hot-spot and reduce performance.

The explanation for the new magnetized design has so far focussed on the impact of increasing the shell thickness. However, the parameter scan in figure \ref{fig:shicethick} suggests that ice thickness must be decreased simultaneously. This is thought to be due to keeping the shock breakout time (when the first shock is released into the DT vapour) the same as the nominal design for N170601. The shock timings are important for setting the implosion adiabat. As the radiation drive is kept the same for all cases in this simplified study, the ice thickness must be decreased with increasing shell thickness to conserve this timing. Indeed, these timings are found in the simulations to be the same for both the unmagnetized optimum design and the magnetized optimum. Future work will also optimize the laser pulse and the resulting shock timing; this added freedom should allow for futher magnetization benefits.

\section{2D Extended-MHD Simulations \label{sec:2D}}

This section takes the improved magnetized design from the previous section (84$\mu m$ shell thickness and 38$\mu m$ ice thickness) and runs full extended-MHD calculations to check the performance. A series of applied magnetic field strengths are used as well as radiation drive asymmetries to correct the implosion shape. Up to 100T magnetic fields are applied, which is much greater than the 26T used so far at the National Ignition Facility \cite{sioDiagnosingPlasmaMagnetization2021}, although new coil designs suggest that this value is feasible. The radiation asymmetry is applied constant in time throughout the whole implosion, which is a simplification of the hohlraum dynamics that are time varying \cite{callahan2018}.

Without tuning the drive symmetry, the 100T field gives a neutron yield of 2.8$\times$10$^{17}$, a factor of 5 below the result given using the 1D code. The shape becomes severely sausaged, particularly once the hot-spot begins to reexpand with substantial $\alpha$-heating (as first noted elsewhere \cite{oneill2025}). However, a radiation asymmetry can be applied to improve the implosion performance \cite{walshMagnetizedICFImplosions2022}; here the P2, P4 and P8 Legendre modes were scanned. Using 1.75\% P2 + 0.5\% P4 + 1.0\% P8 corrects most of the idealized neutron yield, giving 8.9$\times$10$^{17}$, which is a 37\% reduction from the 1D calculation; this corresponds to an energetic yield of 2.51MJ, which would be a clear enhanement over other yields at this laser energy of 1.57MJ. While an initial field of 100T is far higher than possible currently, a 50T field is found to give a yield of 7.9$\times$10$^{17}$, i.e. a large fraction of the 100T yield enhancement. Compared with the optimal 1D unmagnetized design (72$\mu m$ shell + 48$\mu$m ice), the 100T field on the new design (84$\mu m$ shell + 38$\mu$m ice) gives an enhancement of 18.5$\times$.

\begin{figure}
	\centering
	\begin{subfigure}{\linewidth}
		\centering
		\includegraphics[scale=0.5]{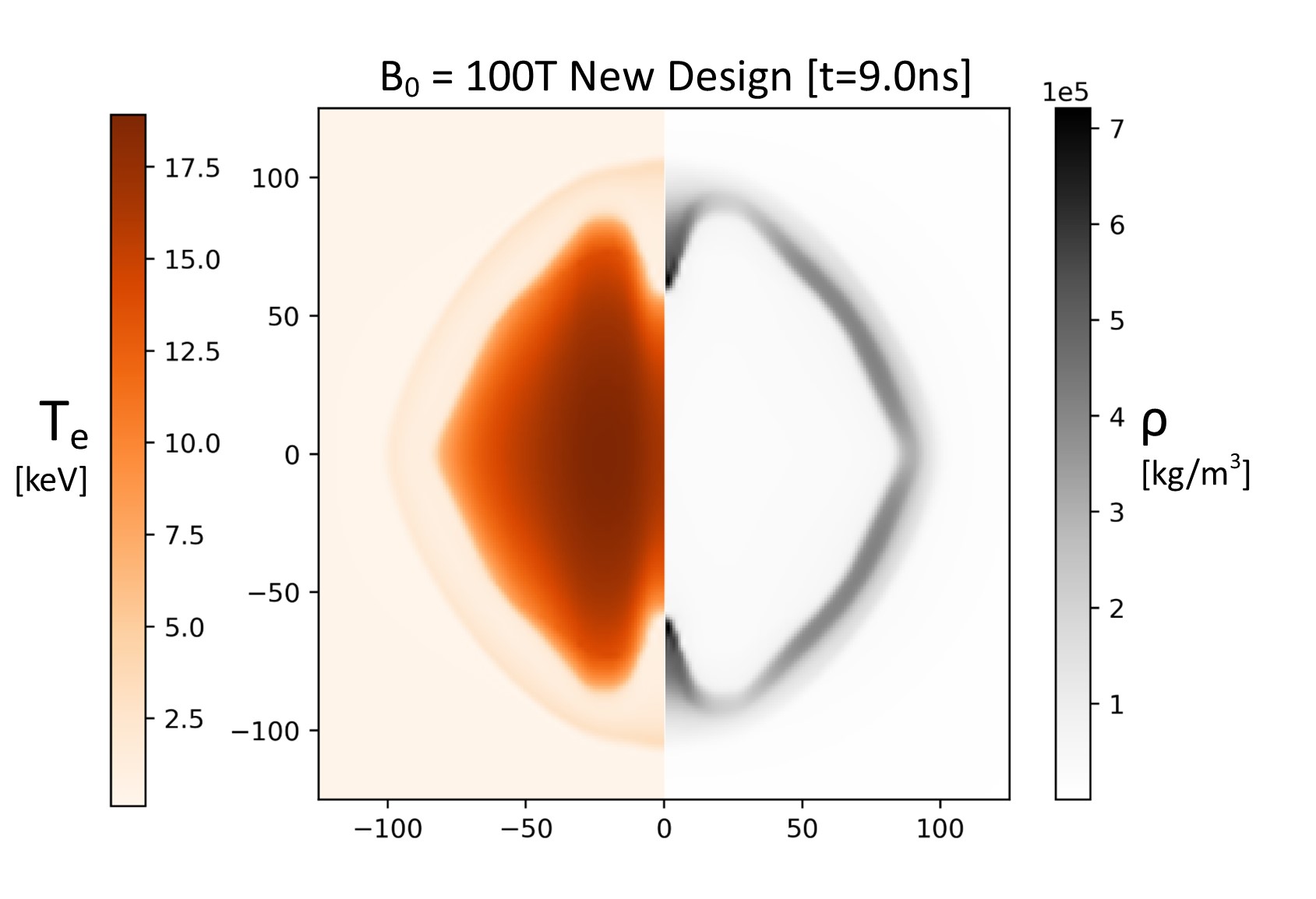}
		\caption{\label{fig:Newdesign_100T}}
	\end{subfigure}
	
	\begin{subfigure}{\linewidth}
		\centering
		\includegraphics[scale=0.5]{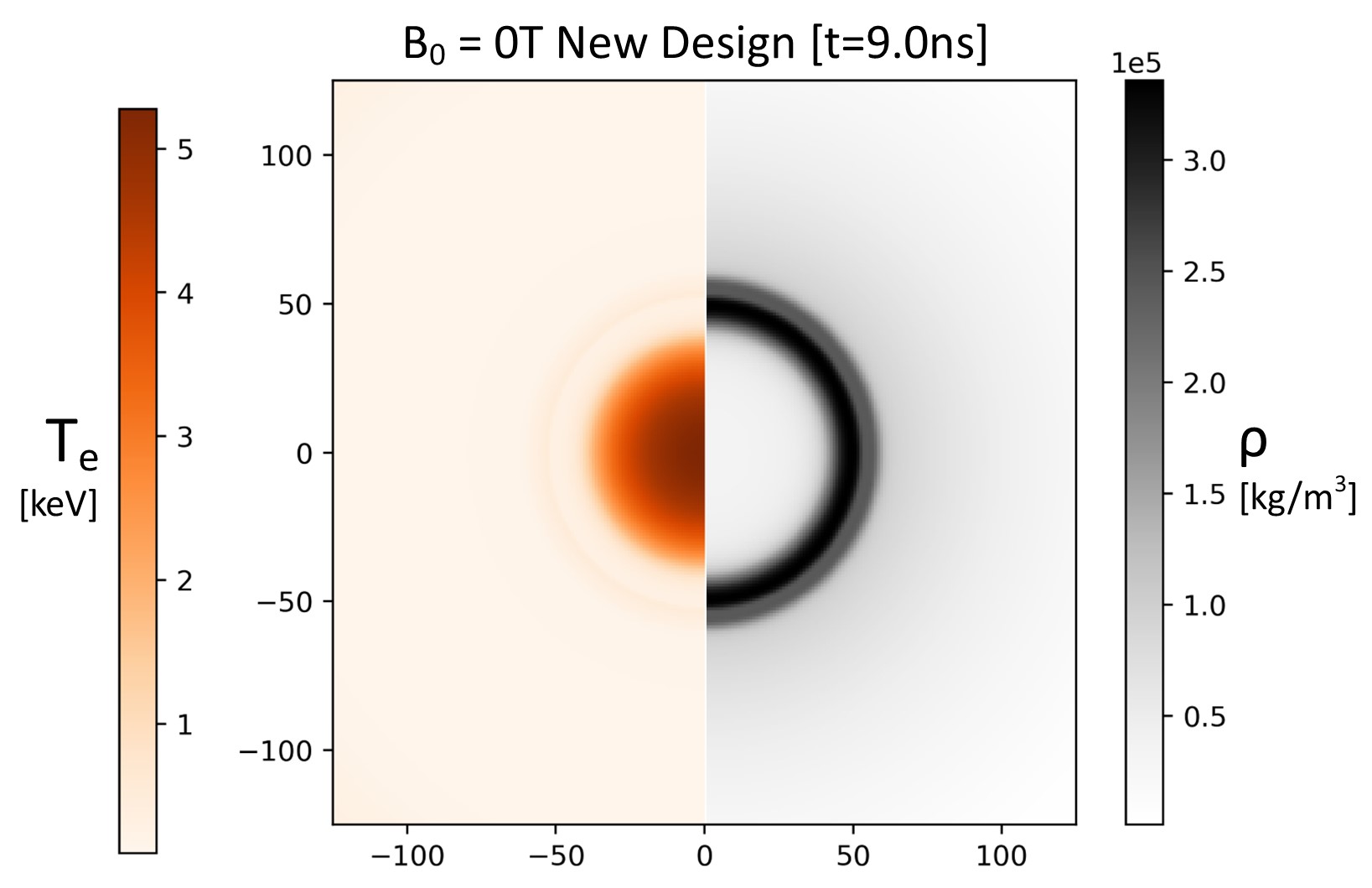}
		\caption{\label{fig:Newdesign_0T}}
	\end{subfigure}
	\caption{ \label{fig:Newdesign} 2D extended-MHD simulations of the new design, showing electron temperature and mass density. (a) is using a 100T initial magnetic field, while (b) is unmagnetized. Both are shown at 9ns, which is the time of peak neutron output for the magnetized case. The unmagnetized case has a neutron yield of 2.6$\times 10^{16}$, while the magnetic field increases that by a factor of 34 to 8.9$\times 10^{17}$. }	
\end{figure}

Figure \ref{fig:Newdesign_100T} shows the hot-spot electron temperature and density for this 100T simulation at peak neutron output (9.0ns). The hot-spot is expanding at this time due to substantial coupling of the $\alpha$-particles back into the fuel, with a radius of almost 100$\mu$m and temperatures in excess of 17keV. A significant polar jet is observed, which has been documented elsewhere for highly-magnetized high-yield implosions \cite{djordjevic2024}. It is possible that the shape could be further refined to increase the yield, but this is left for a later study. 153 2D simulations were already used to scan for this single magnetized target design over P2, P4 and P8.

Figure \ref{fig:Newdesign_0T} shows the unmagnetized equivalent hot-spot for the new design driven symmetrically. The temperature remains relatively low and the fusion yield is only 2.6 $\times$10$^{16}$, a factor of 34 below the shape-tuned 100T simulation. Without the applied magnetic field the design does not ignite.

\section{Conclusion}

This paper finds that the optimum design for a magnetized ICF implosion is different than for an unmagnetized implosion. A design that is below the ignition threshold is taken as the nominal design; 2D extended-MHD simulations anticipate an 18.5-fold increase in neutron yield through the imposition of a magnetic field in combination with increasing shell thickness and decreasing ice thickness. In addition to these calculated benefits, the new design has a lower implosion velocity and more shell mass remaining, which increase ablation-front stability.

The increased performance of the new magnetized design is attributed to increased confinement times. As magnetization enhances the hot-spot temperature, it lowers the constraint on implosion velocity to reach ignition and allows for more shell mass remaining during hot-spot formation. This results in yield amplifications due to $\alpha$-heating in excess of 60.

This simple study uses a fixed radiation drive and only varies the ice and shell thicknesses; this allows for a simple 2D parameter scan of implosion designs. Future work will also optimize the laser drive for the magnetized case and will require a more sophisticated optimization algorithm to explore the higher-dimensional parameter space. The work here demonstrates the utility of using a simplified 1D magnetization model to find new magnetized implosion designs, with comparable performance to more complete 2D extended-MHD simulations.

The current work takes a sub-ignition implosion (N170601) as the baseline design. While this demonstrates that lower laser energies are required to reach ignition for magnetized implosions, it does not investigate the benefit of magnetization for the current high-performing ignition shots that have been conducted on the NIF \cite{abu-shwarebLawsonCriterionIgnition2022,PhysRevLett.132.065102}. This is the next step in redesigning ICF implosions to best account for magnetization.

None of the enclosed work accounts for the benefits of magnetization on mitigating perturbation growth and mix \cite{srinivasanMitigatingEffectMagnetic2013,walshMagnetizedAblativeRayleighTaylor2022}. These are expected to further enhance the yield of implosions with realistic perturbation sources such as the fill-tube and native surface roughness \cite{walshPerturbationModificationsPremagnetisation2019,walshMagnetizedICFImplosions2022}.

\section*{Acknowledgements}
This work was performed under the auspices of the U.S. Department of Energy by Lawrence Livermore National Laboratory under Contract DE-AC52-07NA27344. 

Work supported by LLNL LDRD project 23-ERD-025.

This document was prepared as an account of work sponsored by an agency of the United States government. Neither the United States government nor Lawrence Livermore National Security, LLC, nor any of their employees makes any warranty, expressed or implied, or assumes any legal liability or responsibility for the accuracy, completeness, or usefulness of any information, apparatus, product, or process disclosed, or represents that its use would not infringe privately owned rights. Reference herein to any specific commercial product, process, or service by trade name, trademark, manufacturer, or otherwise does not necessarily constitute or imply its endorsement, recommendation, or favoring by the United States government or Lawrence Livermore National Security, LLC. The views and opinions of authors expressed herein do not necessarily state or reflect those of the United States government or Lawrence Livermore National Security, LLC, and shall not be used for advertising or product endorsement purposes.

The data that support the findings of this study are available from the corresponding author upon reasonable request.

\end{document}